\documentclass[reprint, groupedaddress, superscriptaddress, nobibnotes]{revtex4-2} 
\usepackage{graphicx}
\usepackage[colorlinks, allcolors=blue]{hyperref}

\begin{document}
\title{Ultrafast Laser Induces Macroscopic Symmetry-Breaking of Diamond Color Centers}
\author{Yang Gao}
\author{Qi-Zheng Ji}
\author{Chao-Bo Liu}
\author{Qi Xiao}
\affiliation{Beijing Institute of Spacecraft Environment Engineering,100094 Beijing, P. R. China}
\author{Chao Lian}
\email{Correspondence to: chaolian@iphy.ac.cn}
\affiliation{Beijing National Laboratory for Condensed Matter Physics and Institute of Physics, Chinese Academy of Sciences, Beijing, 100190, P. R. China}
\affiliation{Songshan Lake Materials Laboratory, Dongguan City, Guangdong Province, P. R. China}
\begin{abstract}
The negatively charged nitrogen-vacancy center is a leading quantum platform due to its excellent spin coherence and stable interactions. Understanding its ultrafast dynamics is crucial for quantum applications but presents significant challenges for both experimental characterization and atomic-scale modeling. Here, we employ real-time time-dependent density functional theory to investigate the coupled electron-phonon-spin dynamics in negatively charged nitrogen-vacancy centers. Laser excitation promotes minority-spin electrons within 100~fs, establishing a $C_{3v}$-symmetry breaking charge ordering. Subsequently, ionic motion on the potential energy surface of the excited electrons generates both symmetric oscillations of carbon-nitrogen bonds and dynamic Jahn-Teller distortions with a $C_{3v}$-symmetry breaking. These distortions subsequently induce nonlocal coherent phonons in the diamond lattice, which propagate with the $C_{3v}$-symmetry breaking at the sound velocity ($\sim$2~\AA/fs). Our simulations provide direct time-resolved visualization of these processes, offering novel insights into the microscopic interplay of electrons, phonons, and spins in nitrogen-vacancy centers.
\end{abstract}
\maketitle
~\\\noindent~{Introduction}
\begin{figure}
    \centering
    \includegraphics[width=1.0\linewidth]{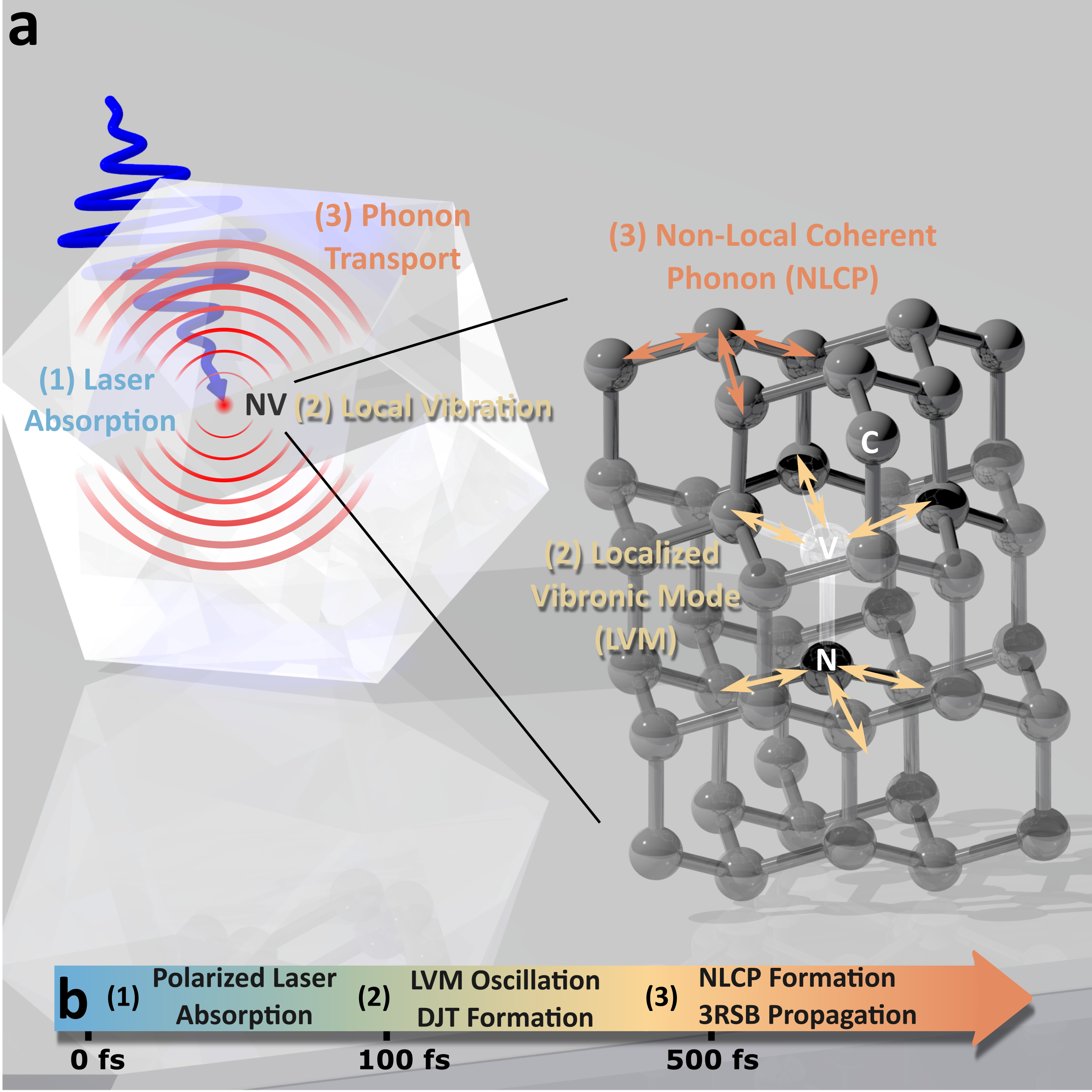}
    \caption{\label{fig:model} Schematic diagrams of laser-excited NV$^{-}$ centers.~{a.} The overall laser-induced dynamics in NV$^{-}$ centers: (1) laser excitation of electrons, (2) induction of local vibrational mode (LVM) vibrations near the NV$^{-}$ centers, and subsequently (3) excitation of nonlocal coherent phonons (NLCP) modes via phonon propagation. ~{b.} Corresponding ultrafast timescales of the processes described in ~{a}.}
\end{figure}

\noindent The negatively charged nitrogen-vacancy (NV$^{-}$) center has become a leading platform in quantum sensing, networking, and computation~\cite{Barry2020Rev.Mod.Phys., Wolfowicz2021NatRevMater, Rovny2024NatRevPhys, Du2024Rev.Mod.Phys.}, due to its exceptional spin coherence time, distinct opto-magnetic and electron-phonon interactions, and remarkable stability over a wide temperature range~\cite{Degen2017Rev.Mod.Phys.}. In these applications, the intricate coupling between light, electronic spin, and phonons, particularly the role of local vibrational modes (LVMs, Figure~\ref{fig:model}~{a}) are crucial in modulating optical properties~\cite{Gali2019Nanophotonics, Gali2023Nanophotonics, Du2024Rev.Mod.Phys.}, while these interactions are entangled in the ground state or weakly perturbed state and thus difficult to understand and manipulate individually.

Pump-probe experiments have revolutionized the study by categorizing the mixed electron-phonon-spin correlations into different timescales in non-equilibrium ultrafast dynamics~\cite{Huxter2013NaturePhys, Ulbricht2016NatCommun, Liu2021Mater.Quantum.Technol., Ulbricht2018Phys.Rev.B, Ulbricht2018Phys.Rev.Ba, Carbery2024NatCommun}. Pioneering work by Huxter \textit{et al.} utilized two-dimensional ultrafast spectroscopy to reveal the dominant role of LVMs in mediating vibrational bath responses~\cite{Huxter2013NaturePhys}. Particularly, the dynamical Jahn-Teller (DJT) distortion has been established as the primary mechanism driving femtosecond-scale depolarization of NV$^{-}$ electronic states~\cite{Ulbricht2016NatCommun}. Recent advances by Carbery \textit{et al.} distinguished the timescales of DJT decay ($\sim$150~fs) and excited-state relaxation (picosecond-scale)~\cite{Carbery2024NatCommun}, while Ichikawa \textit{et al.} demonstrated that a special LVM, dynamical polarons, serve as precursors to long-lived nonlocal coherent phonons (NLCP)~\cite{Ichikawa2024NatCommun}. These results collectively highlight the critical role of phonons in NV$^{-}$ center dynamics from different aspects.

A comprehensive atomic-scale understanding of ultrafast processes in NV$^{-}$ centers is crucial for reconciling these macroscopic spectroscopy into a unified framework. For the ground-state NV$^{-}$ systems, first-principles methodologies have proven instrumental in systematically probing couplings between light and electrons~\cite{Gali2011Phys.StatusSolidiB, Ma2010Phys.Rev.B, Thiering2018Phys.Rev.X, Onizhuk2025Rev.Mod.Phys.}, electrons and phonons~\cite{Deak2010Phys.Rev.B, Gali2011NewJ.Phys., Thiering2017Phys.Rev.B, Ulbricht2018Phys.Rev.B}, and phonons and spins~\cite{Gali2009Phys.Rev.Lett., Ivady2018npjComputMater, Wirtitsch2023Phys.Rev.Res., Kollarics2024Sci.Adv.}. However, modeling non-equilibrium dynamics presents formidable computational challenges: \textit{ab initio} simulations must concurrently resolve spin-orbit-mediated light-electron-phonon-spin interactions, while bridging the six orders of magnitude difference between attosecond electronic time steps and picosecond phonon dynamics. 

Our recent methodological advancements~\cite{Lian2018Adv.TheorySimul., Lian2018TheJournalofChemicalPhysics, Lian2020NatCommun} of the real-time time-dependent density functional theory (RT-TDDFT)~\cite{Runge1984Phys.Rev.Lett., Yabana1996Phys.Rev.B} overcome these limitations through optimized wavefunction propagation schemes that integrate three key components: explicit spin-orbit coupling (SOC) in time-dependent Hamiltonians~\cite{Krieger2015J.Chem.TheoryComput.}, velocity-gauge treatment of electromagnetic fields~\cite{Yabana2012Phys.Rev.B}, and excited-state molecular dynamics capabilities. By applying this advanced RT-TDDFT framework to NV$^{-}$ center systems, we aim to unravel the real-time interplay of electronic, vibrational, and spin degrees of freedom, thereby establishing a new dynamical paradigm for quantum defect research. With the unique atomic-level picture obtained with TDDFT, we propose that the ultrafast combined symmetric oscillations of carbon-nitrogen bonds and DJT distortions propagate through the diamond crystal with the speed of sound and generate a macroscopic coherent symmetry-breaking.

~\\\noindent{Results and Discussion}

\noindent{Photocarrier Generation.}
\begin{figure*}
	\centering
	\includegraphics[width=1.0\linewidth]{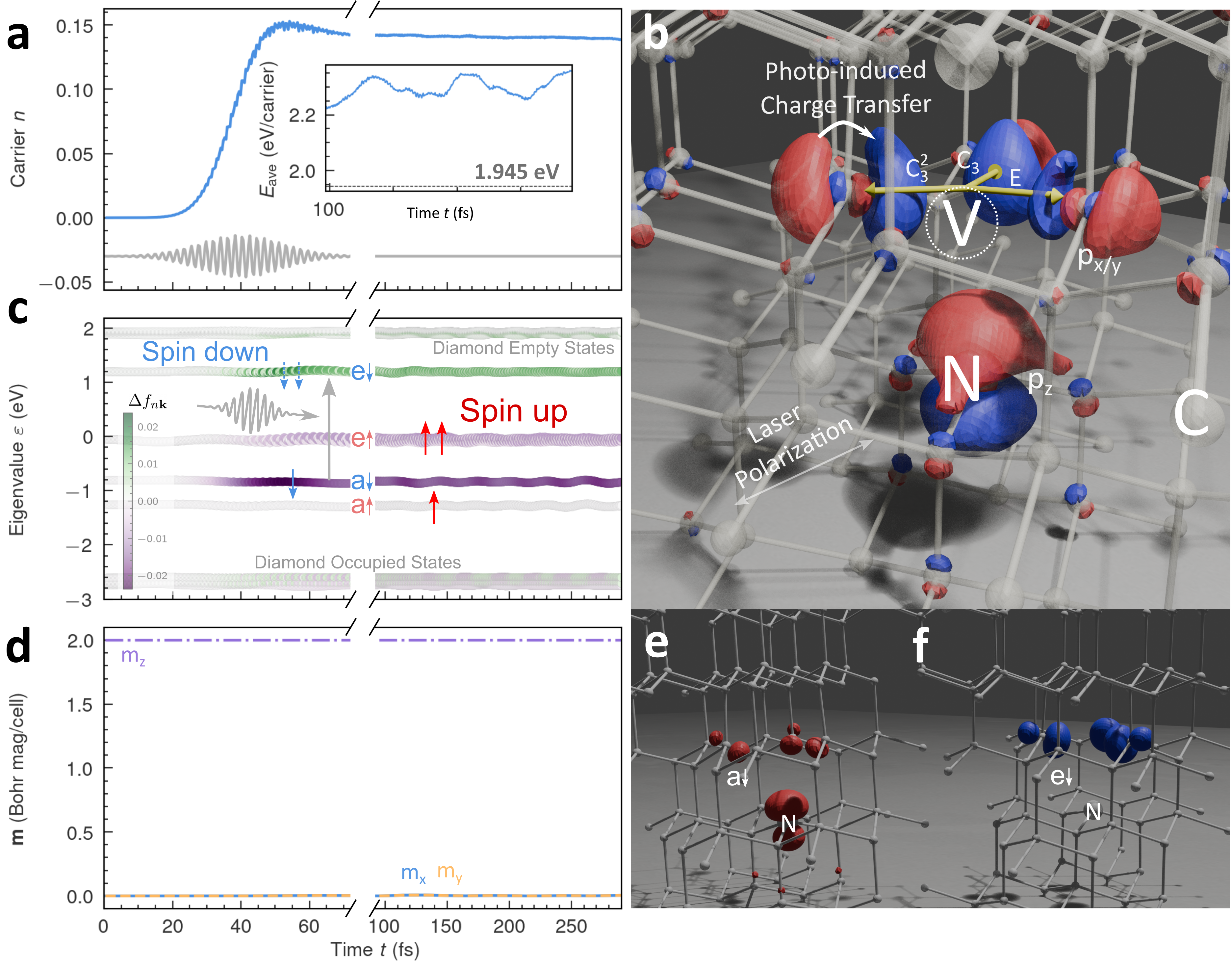}
	\caption{~{Electronic, and spin dynamics in the NV$^{-}$ system.}
    ~{a.}~Number of excited carrier as a function of time. Grey curve: Laser electric field component. Inset: Time-dependent average energy per carrier ($E_\mathrm{ave}$).  
    ~{b.}~Differential electronic density between $t = 100$~fs and $t = 0$, showing charge redistribution (blue: positive $\Delta\rho$; red: negative $\Delta\rho$).  
    ~{c.}~Time-resolved eigenvalue evolution near Fermi energy ($\mathbf{k} = \Gamma$), with color scale ($\Delta f_{n\mathbf{k}}$) indicating electronic population changes.  
    ~{d.}~Evaluation of $m_x$, $m_y$, and $m_z$ as the function of time.
    ~{e} and ~{f}: the wavefunctions of the $a\downarrow$ orbitals and $e\downarrow$ orbitals, respectively. Isosurface level is $5\times10^{-4}$~e/\AA$^3$.}
	\label{fig:electron-spin-dynamics}
\end{figure*}
Under a linearly polarized laser pulse with 100~fs duration, we first analyze the electronic excitation in the NV$^{-}$ center system. The total photoinduced carrier density is quantified as  
\begin{equation}
	\label{eq:photo-carrier}
	n_\mathrm{pc}(t) = \sum_{n\mathbf{k}} \left| f_{n\mathbf{k}}(t) - f_{n\mathbf{k}}(0) \right|,
\end{equation}  
where $ f_{n\mathbf{k}} $ represents the time-dependent electronic population at band $ n $ and crystal momentum $ \mathbf{k} $. During the laser pulse, approximately $ n_\mathrm{pc} = 0.15$ carriers are generated [Figure~\ref{fig:electron-spin-dynamics}~{a}]. The transient energy per carrier is calculated as $E_\mathrm{ave}(t) = [E_\mathrm{KS}(t) - E_\mathrm{KS}(0)]/{n_\mathrm{pc}},$ where $ E_\mathrm{KS}(t)$ denotes the time-dependent Kohn-Sham energy~\cite{Runge1984Phys.Rev.Lett.}. As shown in the inset of Figure~\ref{fig:electron-spin-dynamics}~{a}, $E_\mathrm{ave}$ reaches $ \sim 2.2$~eV/carrier at $t = 100$~fs, consistent with the incident photon energy of $1.945$~eV with energy broadening 0.3 eV. This indicates that the first-order photon absorption dominating at this laser intensity, which is consistent with the experimental setups~\cite{Huxter2013NaturePhys, Ulbricht2016NatCommun, Carbery2024NatCommun, Ichikawa2024NatCommun}. 

The photoinduced charge density redistribution is shown in Figure~\ref{fig:electron-spin-dynamics}~{b}. The spatial asymmetry reveals a $p_z$-orbital-shaped wavefunction localized at the nitrogen (N) atom and three $p_x/p_y$-orbital-like lobes near the vacancy (V) site. This is consistent with the transition from the $ a{\downarrow} $ state to the $ e{\downarrow} $ manifold as shown in Figure~\ref{fig:electron-spin-dynamics}~{e} and ~{f}. Notably, the $ p_x/p_y $-like charge distributions exhibit a $C_{3v}$-symmetry breaking (3RSB): the lobe perpendicular to the laser polarization direction ($ ~{E} $) is suppressed, while the two lobes along the $C_3$ and $C_3^2$ symmetry axes remain equivalent. For $t > 100$~fs, the electron-electron scattering processes promote electrons from the $e{\uparrow}$ states into the conduction bands of bulk diamond.

The spin polarization remains almost unchanged during the subsequent electron-phonon dynamics, as shown in Figure~\ref{fig:electron-spin-dynamics}(d). This agrees well with previous studies that the perpendicular component of the SOC (\(\lambda_\perp\)), which is responsible for spin flipping, is considered negligible in the \(^3E\) triplet~\cite{Doherty2011NewJ.Phys., Maze2011NewJ.Phys.}. This observed spin stability enables the efficient laser initialization of the electronic spin into the \(m_s=0\) ground state~\cite{Barry2020Rev.Mod.Phys.}.  These results reproduce the well-established excitation pathways~\cite{Gali2019Nanophotonics, Barry2020Rev.Mod.Phys., Carbery2024NatCommun} and establishes a rigorous foundation for simulating subsequent coupled electron-phonon processes with full quantum-mechanical fidelity.

~\\\noindent{LVMs and Spin Dynamics.}
\begin{figure*}
	\centering
	\includegraphics[width=1.0\linewidth]{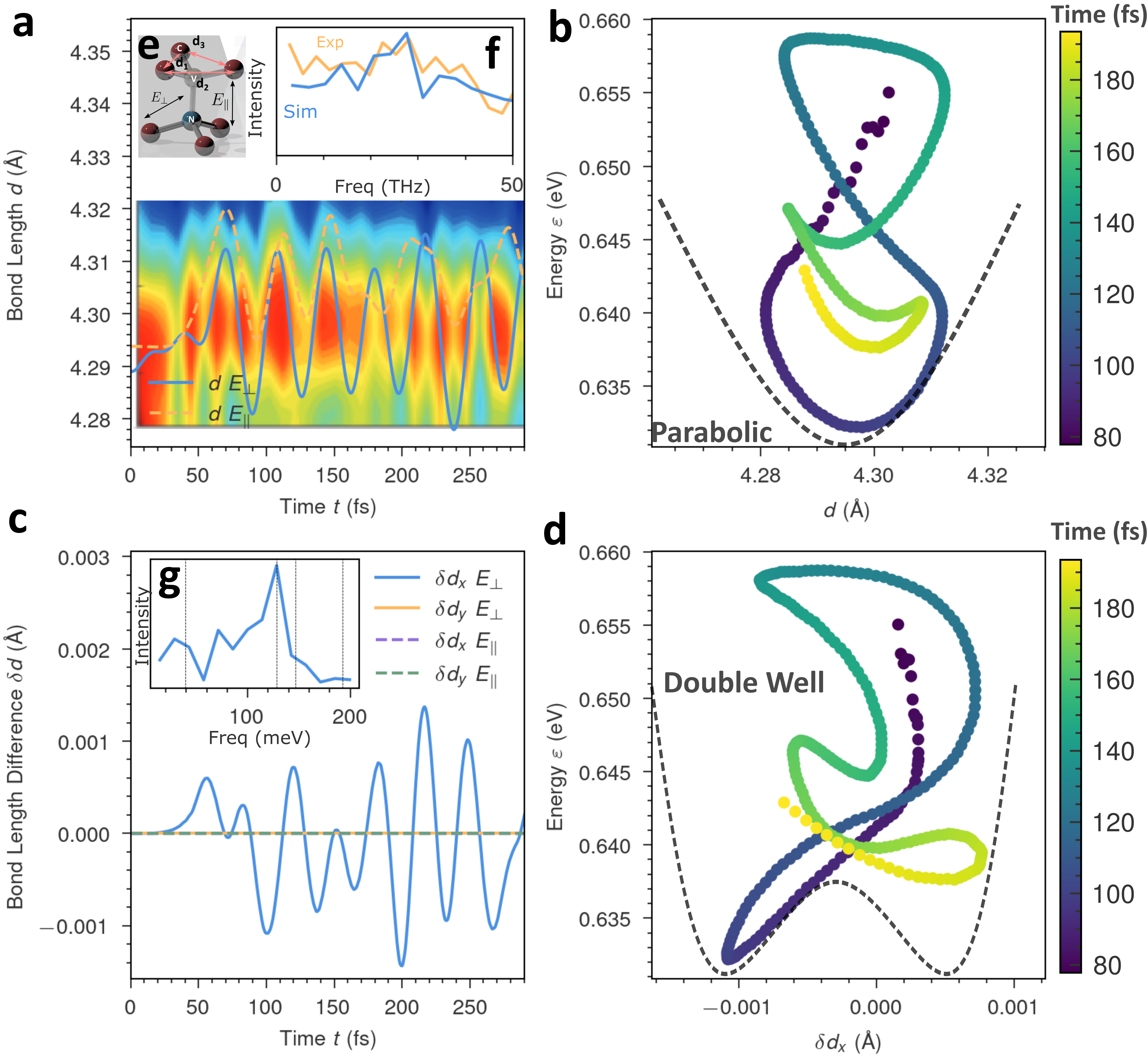}
	\caption{\label{fig:local-vibration}
    ~{Dynamics of the local vibrational mode.}
    ~{a.} Average bond length $d = (d_1 + d_2 + d_3)/\sqrt{3}$ as a function of time, where $d_i$ ($i=1, 2, 3$) are the lengths of the bonds between the nearest carbon atoms and the vacancy (V) site, as illustrated in Inset (left). The solid line represents the result for laser polarization perpendicular to the NV axis ($E_{\perp}$), and the dashed line for polarization parallel to the NV axis ($E_{\parallel}$), as defined in Inset (left). The contour plot is generated using data from Figure 2(f) of Ref.~\cite{Huxter2013NaturePhys}. Inset (right) shows the Fourier transform of $d(t)$ compared with experimental data from~\cite{Huxter2013NaturePhys}.
    ~{b.} Dynamical potential energy surface (PES) as a function of bond length $d$. For comparison, a parabolic PES is indicated by a dashed line. Colors represent different times.
    ~{c.} Differences in bond lengths $\delta d_y$ and $\delta d_x$ as a function of time. Solid and dashed lines denote results for $E_{\perp}$ and $E_{\parallel}$, respectively. Inset shows the Fourier transform of $\delta d_x(t)$ compared with the vibrational frequencies reported in \cite{Abtew2011Phys.Rev.Lett.}.
    ~{d.} Dynamical PES as a function of the bond-length difference $\delta d_x$. A double-well PES is shown as a dashed line for comparison. Colors represent different times.}
\end{figure*}
On longer timescales $>100$~fs, the electron-phonon-spin system exhibits strongly coupled dynamical behavior. First we analyze the LVM dynamics by monitoring the triangle formed by the three carbon atoms around the V site, with lengths \(d_1\), \(d_2\), and \(d_3\), as shown in the left inset of Figure~\ref{fig:local-vibration}~{a}. The average bond length $d = (d_1 + d_2 + d_3)/\sqrt{3}$, is used to describe the symmetric $A_1$ mode and the degenerate $E$ representation is formed by the combinations $\delta d_x = (d_1 + d_2 - 2d_3)/\sqrt{6}, \quad \delta d_y = (d_1 - d_2)/\sqrt{2}$. $d(t)$ oscillations [Figure~\ref{fig:local-vibration}~{a}] with amplitudes of $\sim 0.02$~\AA\ show striking agreement with experimental two-dimensional ultrafast spectra~\cite{Huxter2013NaturePhys} in both time and frequency domains, confirming that the LVMs directly modulate the optical properties of NV$^{-}$ centers.

More importantly, $\delta d_x$ exhibits a oscillation about $1\times10^{-3}$~\AA\ [Figure~\ref{fig:local-vibration}~{c}], which is the microscopic characteristic of DJT distortion~\cite{Abtew2011Phys.Rev.Lett.}. In comparison, when the laser polarization is aligned parallel to the NV center axis [left inset of Figure~\ref{fig:local-vibration}~{a}], the 3RSB is eliminated while the oscillations of LVMs remain qualitatively similar, as shown in Figure~\ref{fig:local-vibration}~{a} and ~{c}. This indicates that the DJT distortion is a direct result of 3RSB charge transfer. The photo-induced charge transfer is weaker along the $\mathbf{E}$-axis compared to that along $C_3$ and $C_3^2$ directions (Figure~\ref{fig:electron-spin-dynamics}~{b}), which causes a 3RSB in the photo-induced ion forces. The peaks in the Fourier transform of \(\Delta d_y\) spectrum, shown in the inset of Figure~\ref{fig:local-vibration}~{c}, are consistent with the E symmetry vibronic levels reported in Ref. \cite{Abtew2011Phys.Rev.Lett.}, providing another characteristic fingerprint for comparison with \(\Delta\)-SCF calculations~\cite{Abtew2011Phys.Rev.Lett.}.

We analyze the bond dynamics with the potential energy surface (PES) by analyze the $E_\mathrm{KS}(t)$ along these two coordinates $d$ and $\delta d_x$, i.e. $E_\mathrm{KS}(d)$ and $E_\mathrm{KS}(\delta d_x)$.$E_\mathrm{tot}(d)$ shows a parabolic PES with a minimum at the excited-state equilibrium bond length (Figure~\ref{fig:local-vibration}~{b}), corresponding to overall oscillations, while $ E_\mathrm{tot}(\delta d_x) $ reveals a double-well PES (Figure~\ref{fig:local-vibration}~{d}) characteristic of the DJT distortion.

This PES evolution has great influence on the structural response, similar to the PES modification predicted~\cite{Lian2020NatCommun} and the non-linear pump-probe signal observed in TiSe$_2$~\cite{Burian2021Phys.Rev.Research}. This directly resolve these competing effects with a critical advance over previous excited-state PES studies relying on perturbative methods like $\Delta$SCF~\cite{Abtew2011Phys.Rev.Lett.}.

~\\\noindent{Phonon propagation.}
\begin{figure*}
	\centering
	\includegraphics[width=1.0\linewidth]{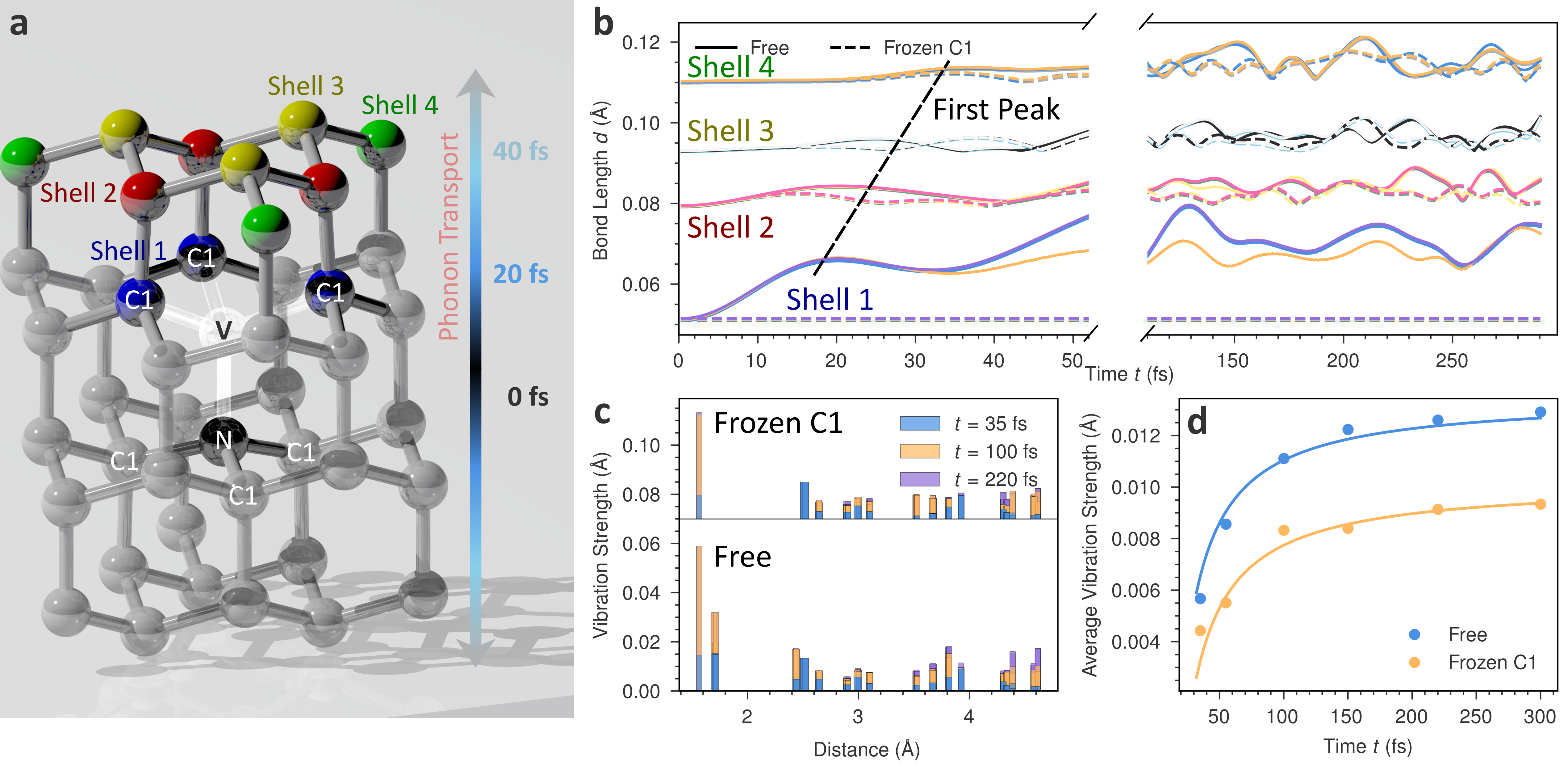}
	\caption{\label{fig:phonon-propagation}
    ~{Macroscopic propagation of the $C_{3v}$-symmetry breaking.}
    ~{a.} Schematic of carbon atoms around the NV$^{-}$ center in different shells and phonon transport. The atoms are categorized into different shells according to their distances to the vacancy (V) site and are marked in different colors. C1 atoms are those nearest to the NV$^{-}$.  
    ~{b.} Displacement $d(t)$ of carbon atoms in different shells as a function of time. The curve for each shell is shifted upward proportionally to its distance from the V site. The solid lines indicate the case with all atoms free, while the dashed lines indicate the case where C1 atoms are fixed.  
    ~{c.} Maximum displacement amplitude as a function of distance to the V site at different final times $t_f$ (35, 100, and 400 fs). The upper panel shows results with C1 atoms fixed, and the lower panel shows results with all atoms free.  
    ~{d.} Average maximum displacement of carbon atoms as a function of $t_f$. Blue lines and symbols correspond to the case with all atoms free, and yellow lines and symbols correspond to the case with C1 atoms fixed.}
    \end{figure*}
Beyond local symmetry breaking, the DJT distortion induces macroscopic NLCPs with 3RSB in diamond. We analyze the oscillation amplitudes $\Delta \tau_i(t) = \mathbf{\tau}_i(t) - \mathbf{\tau}_i(0)$ of the C atoms in successive coordination shells (1–4) around the V site (Figure~\ref{fig:phonon-propagation}~{a}), where $\tau_i(t)$ is the position of the $i$-th C atom. Figure~\ref{fig:phonon-propagation}~{b} shows that during 0–50~fs, $\Delta \tau_i(t)$ values are degenerate for all three C atoms in each shell, while over a longer period (50–150~fs), the atoms along the $C_3$ and $C_3^2$ directions retain degenerate amplitudes, and the atom along the $\mathbf{E}$-axis exhibits a reduced amplitude, indicating the emergence of 3RSB in NLCPs besides the LVMs.  

This 3RSB NLCPs propagates radially, evidenced by a delayed onset and reduced amplitudes in outer shells. The 40~fs delay between the NV$^{-}$ center and shell 4 agree well with the 50~fs NLCP formation time proposed in~\cite{Ichikawa2024NatCommun}. Quantifying displacements via the maximum amplitude $\max\{\Delta \tau_{i}(t_f)\}$ for $t < t_f$ (Figure~\ref{fig:phonon-propagation}~{c}), we find that only LVMs are excited at $t = 35$~fs, while oscillation amplitudes saturate for C atoms in shells 1–3 at $t = 100$~fs.
The average maximum amplitude over the whole cell at different $t_f$ (Figure~\ref{fig:phonon-propagation}~{d}) is evaluated as  
\begin{equation}  
\overline{\max\{\Delta \tau_{i}(t_f)\}} = \frac{1}{N_\mathrm{atom}}\sum_{i=1}^{N_\mathrm{atom}} \max\{\Delta \tau_{i}(t_f)\},  
\end{equation}  
where $N_\mathrm{atom}$ is the total number of atoms in the calculated region. We found that the $\overline{\max\{\Delta \tau_{i}(t_f)\}}$ can be well described with a diffusion-like relation:
\begin{equation} 
\label{eq:diffusion_mode}
\overline{\max\{\Delta \tau_{i}(t_f)\}} = A \exp\left(-\frac{1}{4Dt_f}\right) + C,  
\end{equation}  
with A = 0.20~\AA, C = -0.19~\AA, and D = 0.25~\AA/fs. The diffusion speed D is close to the experimental sound velocity of diamond (0.19~\AA/fs)~\cite{Wang2004MaterialsChemistryandPhysics}, confirming the diffusion nature of the NLCP. Thus, we demonstrate that the DJT distortions trigger the macroscopic symmetry-breaking phonon transport.

The origin of the NLCP is the electric field generated by the excited-state electron distribution [Figure~\ref{fig:electron-spin-dynamics}]. We carry a calculation with six carbon atoms near the NV center fixed, i.e., the C1 atoms in Figure~\ref{fig:phonon-propagation}~{a}. As compared in Figure \ref{fig:phonon-propagation}~{b}, it can be observed that not only is the onset of vibrations delayed for carbon atoms in different shells away from the NV center, but the vibrational amplitude also remains very small within the first 50 fs. Utilizing the previously described diffusion model \ref{eq:diffusion_mode} to characterize this propagation process, we obtain a new set of parameters with A = 0.25~\AA, C = -0.24~\AA, and D = 0.25~\AA/fs. This indicates that diffusion mechanism is similar and the propagation speed remains the sound velocity of diamond.

~\\\noindent{Summary and Discussion}
In this work, we have elucidated the ultrafast coupled dynamics of electrons, phonons, and spins in the NV$^{-}$ center following photoexcitation by leveraging RT-TDDFT. Our simulations directly capture a multi-stage dynamical sequence: Laser excitation initially promotes minority-spin electrons within 100~fs, establishing a charge-ordered state that breaks the $C_{3v}$ symmetry. This electronic rearrangement subsequently drives ionic motion, generating both symmetric bond oscillations and DJT with 3RSB. These local distortions, in turn, launch NLCP wavepackets that propagate through the diamond lattice at the sound velocity, carrying the 3RSB signature. 

The key finding of this study is the direct, time-resolved visualization of the process of the laser-induced 3RSB transfer, from the electronic excitation into specific lattice vibrations and propagation. These insights provide a novel fundamental understanding of the NV$^{-}$ center's photophysics and suggest potential avenues for coherent control of its quantum state via tailored phononic environments, with implications for quantum information science.

~\\\noindent{Methods}

\noindent Based on our previous efforts in developing the time-dependent \textit{ab initio} package (TDAP)~\cite{Lian2020NatCommun}, alongside the velocity-gauge field algorithm~\cite{Lian2018Adv.TheorySimul., Lian2018TheJournalofChemicalPhysics} and molecular dynamics~\cite{Meng2008TheJournalofChemicalPhysics, Lian2018TheJournalofChemicalPhysics}, we implemented the TDDFT algorithm~\cite{Wang2015Phys.Rev.Lett.} within the plane-wave code \textsf{Quantum Espresso}~\cite{Giannozzi2009J.Phys.:Condens.Matter, Giannozzi2017J.Phys.:Condens.Matter}. For details on the velocity-gauge electric field, the efficient evolution method, choice of the XC functionals, and the energy cutoff/~{k}-points convergence, please refer to the Supplementary Information~\cite{Supplementary_Information} of Ref.~\cite{Lian2020NatCommun}. We used the Perdew-Burke-Ernzerhof (PBE) exchange-correlation (XC) functional~\cite{Perdew1996Phys.Rev.Lett.} in both DFT and TDDFT calculations. The core electrons and nuclei were described using full-relativistic, optimized norm-conserving Vanderbilt pseudopotential (ONVP)~\cite{Hamann2013Phys.Rev.B} from the \textsf{PseudoDojo} pseudopotentials library~\cite{vanSetten2018ComputerPhysicsCommunications}. The plane-wave energy cutoff was set to 80~Ry. The NV system is modeled as a 3$\times$3$\times$3 supercell of diamond with one NV center, containing 53 atoms (52 C atoms and 1 N atom). Brillouin zone was sampled using Monkhorst-Pack scheme~\cite{Monkhorst1976Phys.Rev.B} with a $3\times3\times3$ k-point mesh. Band unfolding techniques were utilized to generate the effective band structures (EBS)~\cite{Popescu2012Phys.Rev.B, Lian2017Phys.Rev.B} with \textsf{unfold-x} code~\cite{Pacile2021AppliedPhysicsLetters}. The electron timestep $\delta t$ is $4\times10^{-4}$~a.u.=$0.194$~attosecond and the ion timestep $\Delta t$ is $0.194$~fs. Structural figures are generated with VESTA~\cite{Momma2011JApplCrystallogr}.

The Gaussian-type laser pulse is utilized
\begin{equation}
\label{eq:GaussianWave}
\mathbf{E}(t)={E}_0\hat{\mathbf{x}}\cos\left(\omega t \right)\exp\left[-\frac{(t-t_0)^2}{2\sigma^2}\right],
\end{equation}
where $E_0 = 0.03$ V/$\mathrm{\AA}$ is the peak field strength, $\omega = 1.945$ eV is the laser frequency, $t_0 = 40$~fs is the Gaussian peak time, and $\sigma = 12$ fs is the Gaussian pulse width. We note here that the $\omega = 1.945$ laser frequency used in the simulations does not physically correspond to the experimental ZPL excitation energy due to the well-known bandgap underestimation of the PBE functional. The choice of $\omega = 1.945$ here is to be resonant with the calculated vertical excitation energy on the PBE energy scale. Please see the Supplementary Information for detailed discussion. $\hat{\mathbf{x}}$ denotes that the laser pulse is linearly polarized along the $x$ direction. We have tested on different laser intensity ${E}_0$ from $0.01$~V/\AA\ to $0.06$~V/\AA\ and find that the results are similar with different amplitudes.

~\\\noindent{Data availability}

\noindent The key data underpinning the conclusions of this work are included in the article. All raw data relevant to the findings of this study are available from the corresponding author (C.L.) upon request.

~\\\noindent{Code availability}

\noindent Part of the computational codes that were used to generate the data is presented \cite{Lian2026} and that the full code can be provided by the corresponding author (C.L.) upon request.

~\\~\\\noindent{Acknowledgment}

\noindent C.L. thanks \'{A}d\'{a}m Gali for the helpful discussion. We acknowledge the financial supports from National Key Research and Development Program of China, Grant No.~2024YFA1408603 and CAS Project for Young Scientists in Basic Research under grant No.~YSBR-120. Computational resources were provided by Xi'an Buhan and China HPC Platform.

~\\\noindent{Author contributions}

\noindent C.L. proposed the study, supervised and obtained funding to support the project. Y.G. and C.L. performed the calculations and data analysis. Y.G. and C.L. drafted the initial paper, and all authors contributed to the writing of the manuscript.

~\\\noindent{Competing interests}

\noindent The authors declare no competing interests.

\end{document}